# The effect of the process of star formation on the temporal evolution of the WR and O-type star populations in the Solar Neighbourhood

**D. Vanbeveren and E. De Donder**

Astrophysical Institute,
Vrije Universiteit Brussel, Pleinlaan 2, 1050 Brussels, Belgium

**Abstract** A comparison between the observed O-type star and the WR star populations and the theoretically predicted ones depends on the effects of stellar wind mass loss during various phases and rotation on stellar evolution and, last not least, on the effects of binaries. Obviously, both populations depend on the massive star formation rate. In the present paper we show that the rate in the Solar Neighbourhood is fluctuating on a timescale which is smaller than the evolutionary timescale of a massive star. We demonstrate that this fluctuating behaviour affects both population enormously, in such a way that a comparison between observations and theoretical prediction to test stellar evolutionary models is ambiguous.

**Introduction**

A comparison between the theoretically predicted populations of Wolf-Rayet (WR) stars and O-type stars and the observed ones in the Solar Neighbourhood (SoNe) has been performed frequently in the past in order to test evolutionary models of massive stars. Probably the most complete study was published in Vanbeveren et al. (1998). In this study we accounted in detail for the evolution of massive single stars and of massive close binaries. Below we list the processes which were included and the assumptions which were used:

- massive star formation was continuous and constant during the last 10 Myr

- the formation timescale of a massive star is short compared to its O-type lifetime (= the lifetime during which a massive star will be observed as an O-type star)

- we need an unambiguous definition of a WR star which enables us to decide when a stellar model calculated with an evolutionary code can be classified as a WR star. WR stars are hydrogen deficient stars (Xatm $\leq$ 0.4). However, at least for binaries this definition is insufficient since most of the primaries in interacting binaries lose almost all of their hydrogen rich layers due to Roche lobe overflow (RLOF) and start core helium burning as hydrogen deficient stars. Therefore an extra condition is needed. Using the observed minimum mass of WR stars in binaries seems logical. Accounting for observational uncertainty this minimum mass $\approx 5 - 8\ M_o$ for the SoNe; from observational point of view it is not known whether this value depends on metallicity or not.

- we used evolutionary calculations with moderate amount of convective core overshooting during core hydrogen burning (0.2 times the pressure scale height evaluated at the Schwarzschild boundary of the convective core, as promoted in Schaller et al., 1992)

- possible effects of rotation were not included (we add a discussion below)

- evolutionary calculations depend on the stellar wind mass loss rate formalisms which are implemented in the computer code. Of particular importance are the formalisms which apply for Luminous Blue Variables (LBV) with an initial mass larger than 40 $M_o$, the mass loss during the Red Supergiant (RSG) phase and during the WR phase. The reader should pay attention at the mass loss rate formalism that we adopted during the RSG phase of a single star which is significantly different (we think that it is also more realistic) from the formalism preferred by other groups, e.g. the Geneva group. Notice that the RSG formalism affects in a critical way the predicted WR population of single stars. Furthermore, already in 1998 we proposed a WR stellar wind mass loss formalism which predicts similar rates as the formalisms recently used by most of the other groups

- binaries with a primary mass larger than 40 $M_o$ are assumed to evolve according to the 'LBV scenario' (see Vanbeveren, 1991) which means that for population number synthesis (PNS) predictions, it hardly matters whether a star with initial mass larger than 40 $M_o$ is the primary of a binary or not

- In order to simulate the overall effects of binaries,

  - we calculate the percentage of binaries that merge. We adopt a model for the further evolution of mergers. However, notice that the evolution of mergers has never been studied in detail

  - for the binaries that may evolve through a classical RLOF phase, we consider the evolutionary effects of mass transfer, mass accretion and possible mass loss from the system

  - for the binaries that evolve through a common envelope phase, we use the prescription of Webbink (1984) and the update of de Kool (1990)

- the effects of the supernova (SN) explosion on binary parameters are treated in detail.

Our model is able to reproduce the observed WC/WN number ratio in the SoNe, it predicts a very small number of WR stars with a compact companion corresponding to observations. Depending on the adopted minimum mass for a WR star, the WR/O number ratio ranges between 0,03 (minimum mass = 8 $M_o$) and 0,07 (minimum mass = 5 $M_o$), i.e. compared to the observations (WR/O ~ 0,1) the predicted number is rather small, especially when it is assumed that the minimum WR mass = 8 $M_o$ (or larger of course). The most important conclusion resulting from our study was that including binaries in PNS studies of O type stars and WR stars is essential; not including them may give academic valuable results but they may be far from reality.

Recently, Meynet and Maeder (2003) investigated the effects of rotation on massive star evolution and on the predicted O-type and WR star population in the SoNe. As usual, they entirely denied the effects of binaries but we find it beyond the scope of the present paper in order to restart a discussion on this matter. We leave it to the interested reader in order to make up her/his mind about this

To illustrate the effects of rotation on the WR and O-type star population, Meynet and Maeder (2003) compare theoretical predictions when evolutionary results (of single stars only) are used without rotation with those when it is assumed that on the average stars start their life with a rotational velocity $v_e$ = 300 km/s (corresponding to an average rotational velocity for O-type stars $<v_e>$ ~ 220-240 km/s and $<v_e>$ ~ 240 km/s for early B-type stars). With the O-type star data of Penny (1996) and the early B-type star data (Be, Bnn_n included) of the Bright Star Catalogue, Vanbeveren et al. (1998) computed the expected $v_e$-distribution for both types of stars. Both are shown in Figure 1. As can be noticed, the distributions are highly asymmetrical with a tail extending towards very high rotational velocities. This means that it is very dangerous in order to use evolutionary computations with some average rotational velocity in order to make PNS predictions. By inspecting both figures, it follows that 80% (respectively 70% ) of the O-type stars (resp. early B-type stars) belong to the symmetrical part of the distribution with an average $\langle v_e \rangle$ ~ 108 km/s (resp. ~ 80 km/s). From the evolutionary computations of the Geneva group, it follows that the effects of rotation for this group of stars is similar to the effects of moderate amount of convective core overshooting as it was introduced by Schaller et al. (1992) and can not be simulated by the calculations which correspond to an average initial $v_e$ = 300 km/s.

However, a comparison between the predicted and observed O-type star and WR star populations in the SoNe may become very uncertain if we account for recent and realistic models of star formation in galaxies. This will be the topic of the next two sections.

**2. Massive star formation in the Solar Neighbourhood**

Locally, star formation in the Galaxy is regulated by the balance between cooling due to gas accretion onto the disk, which enhances star formation, and gas heating due to the feedback of massive stars via stellar winds and SN explosions which reduces the star formation efficiency. This means that the star formation rate is a function of the local gravitational potential and thus also depends on the total surface mass density (Talbot and Arnett, 1975; Dopita and Ryder, 1994; Edmunds and Pagel, 1984; Ryder, 1995). A star formation stop at surface gas densities lower than ~ 7 $M_o/pc^2$ has been suggested by observational studies on massive star formation in external galaxies (Kennicut, 1989; Chamcham et al., 1994; Gratton et al., 2000; de Blok and McGaugh, 1996). At these low densities the gas is theoretically expected to be stable against density condensations which are the seeds for star formation. This means that when locally the surface gas density is close to the threshold of ~ 7 $M_o/pc^2$ star formation may go on and off. It has been demonstrated by Chiappini et al. (1997) that the threshold causes a gap in the star formation rate during the evolution of the Galaxy between the end of the halo, thick disk phase and the beginning of the thin disk phase, which naturally explains the observed steep increase of [Fe/O] and [Fe/Mg] at a certain value of [O/H] and [Mg/H] (Gratton et al., 1996).

In order to study the overall evolution of our Galaxy, we combined our PNS model which include binaries, the model of star formation proposed by Talbot and Arnett (1975), a model to calculate the formation and the evolution of the structure of the Galaxy, and a detailed set of single star and binary evolutionary computation which enable us to follow the chemistry and the gas content: all together the Galactic Chemical Evolutionary Model (Galactic CEM). It was described in a series of papers (De Donder and Vanbeveren, 2002, 2003a, 2003b, 2003c) and in De Donder (2004). Figure 2a illustrates the expected temporal evolution of the star formation rate. Notice that a similar behaviour was obtained by Chiappini et al. (1997) although binaries were not explicitly considered in their CEM. Figure 2b shows a blow-up of the SoNe at present. The figures illustrates the following conclusion:

*the surface gas density in the SoNe at present fluctuates around the threshold of 7 $M_o/pc^2$ indicating that during the last few Gyrs the massive star formation rate was not constant and varies within a period which is smaller than the typical evolutionary timescale of a massive star.*

It is obvious to understand that a fluctuating massive star formation rate may have important consequences on the theoretically expected temporal variation of the O-type star and WR star populations of the SoNe. This will be discussed in the next section.

**3. The temporal evolution of the massive star population in the SoNe.**

With our Galactic CEM it is fairly straightforward to calculate the temporal evolution in the SoNe of the WR/O number ratio (it is easy to understand that the temporal evolution of the WC/WN number ratio and the WR+OB binaries/WR ratio are less affected by a variable massive star formation rate). Figures 3 shows the results for the standard CEM model defined in De Donder and Vanbeveren (2002). Although the quantitative values depend on a number of parameters in the CEM, the overall fluctuating behaviour of the  ratio is very similar in all cases. The blow-up of the present day SoNe indicates that the WR/O number ratio fluctuates by a factor 10 in a time interval of at most a few million years. The figure then illustrate the following overall conclusion:

*It is probable that the massive star formation rate at present in the Solar Neighbourhood is not constant and fluctuates on a timescale which is shorter than the typical evolutionary timescale of a massive star. This fluctuating behaviour affects significantly the O-type star and WR star population predicted by theory. It is easy in order to recover the observed WR/O ratio but this possible fluctuation may also be the reason why a comparison between observations in the SoNe and theoretical prediction to test stellar evolutionary models is ambiguous.*

## 4. Final remarks

The star formation model for the SoNe used in the present paper is a uniform density model. This is probably not correct. Even if the average surface density in the SoNe is smaller than the star formation threshold value, it may be larger locally and star formation continues there. In our model star formation suddenly stops but the transition from high to low star formation in the SoNe is probably much smoother. This should also smooth the temporal behaviour of the WR/O number ratio but the overall conclusion (illustrated in Fig. 4) will remain.

Is there any observational evidence that the massive star formation rate is variable? One test may be to collect a large data set of massive stars with well defined luminosities and effective temperatures, plot them in the HR diagram together with predicted time-isochrones, and count the number of stars within the isochrones. This has been done by Vanbeveren (1990) using the data set of Humphreys and McElroy (1984) and it was concluded that the massive star formation rate in the SoNe is declining. We are obviously aware of the problems of the distance scale of massive stars used by Humphreys and McElroy (1984), but until a better data set is available, the study of Vanbeveren (1990) is still indicative.

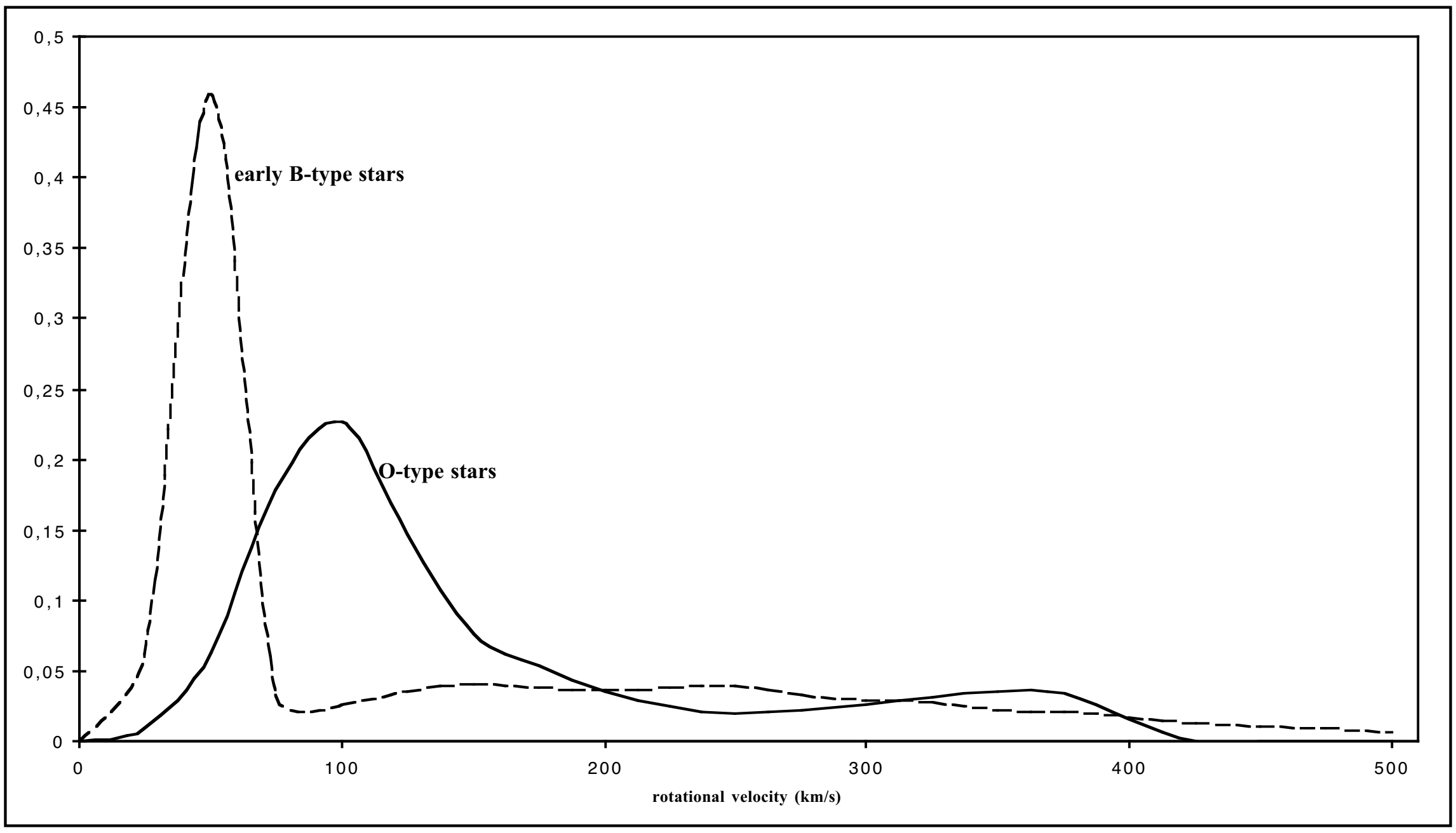

**Figure 1:** The observed distribution of rotational velocities of O-type stars and of early B-type stars.

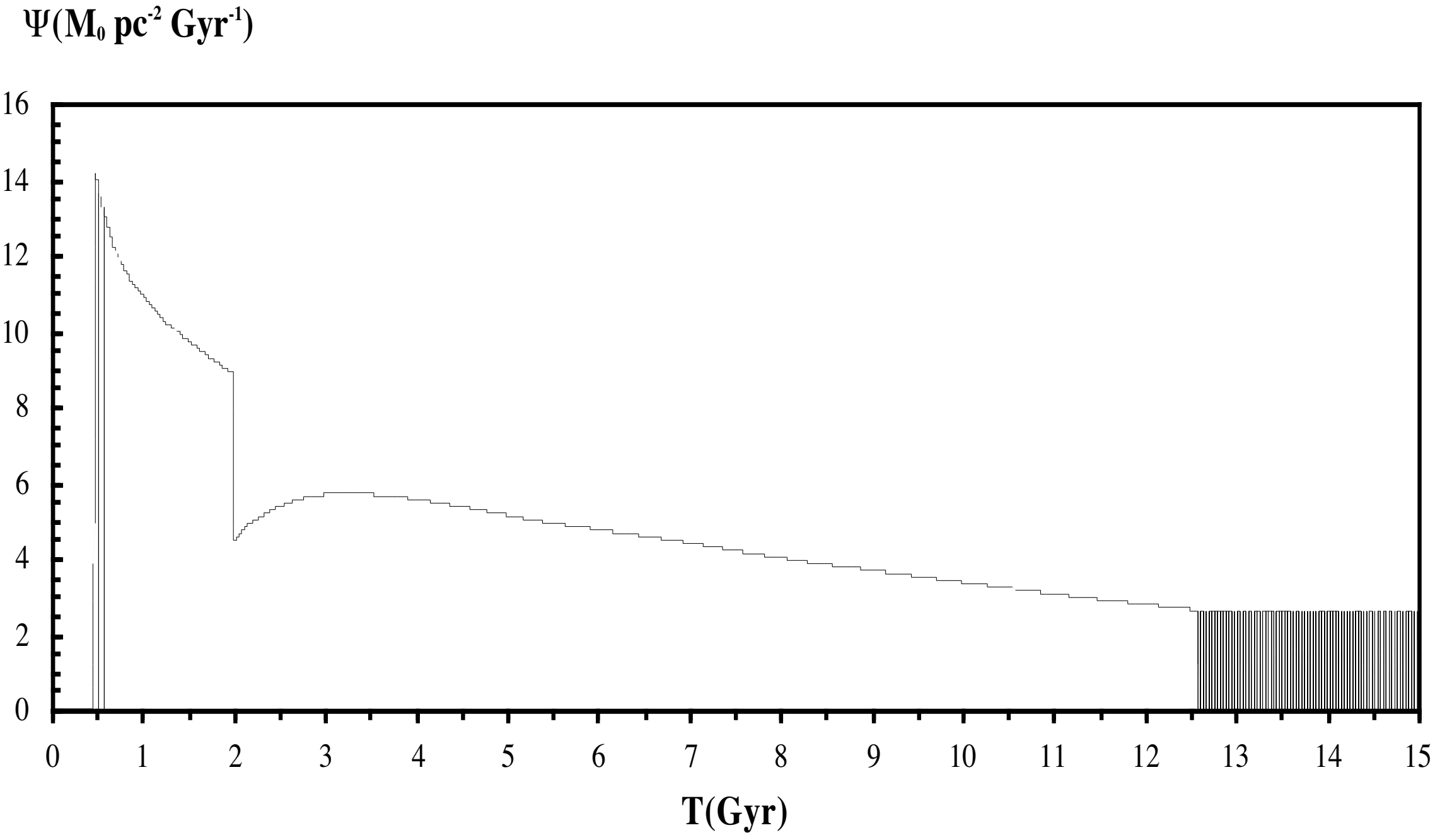

**Figure 2a:** The temporal evolution of the star formation rate in the Solar neighbourhood predicted by Galactic Chemical Evolutionary Models.

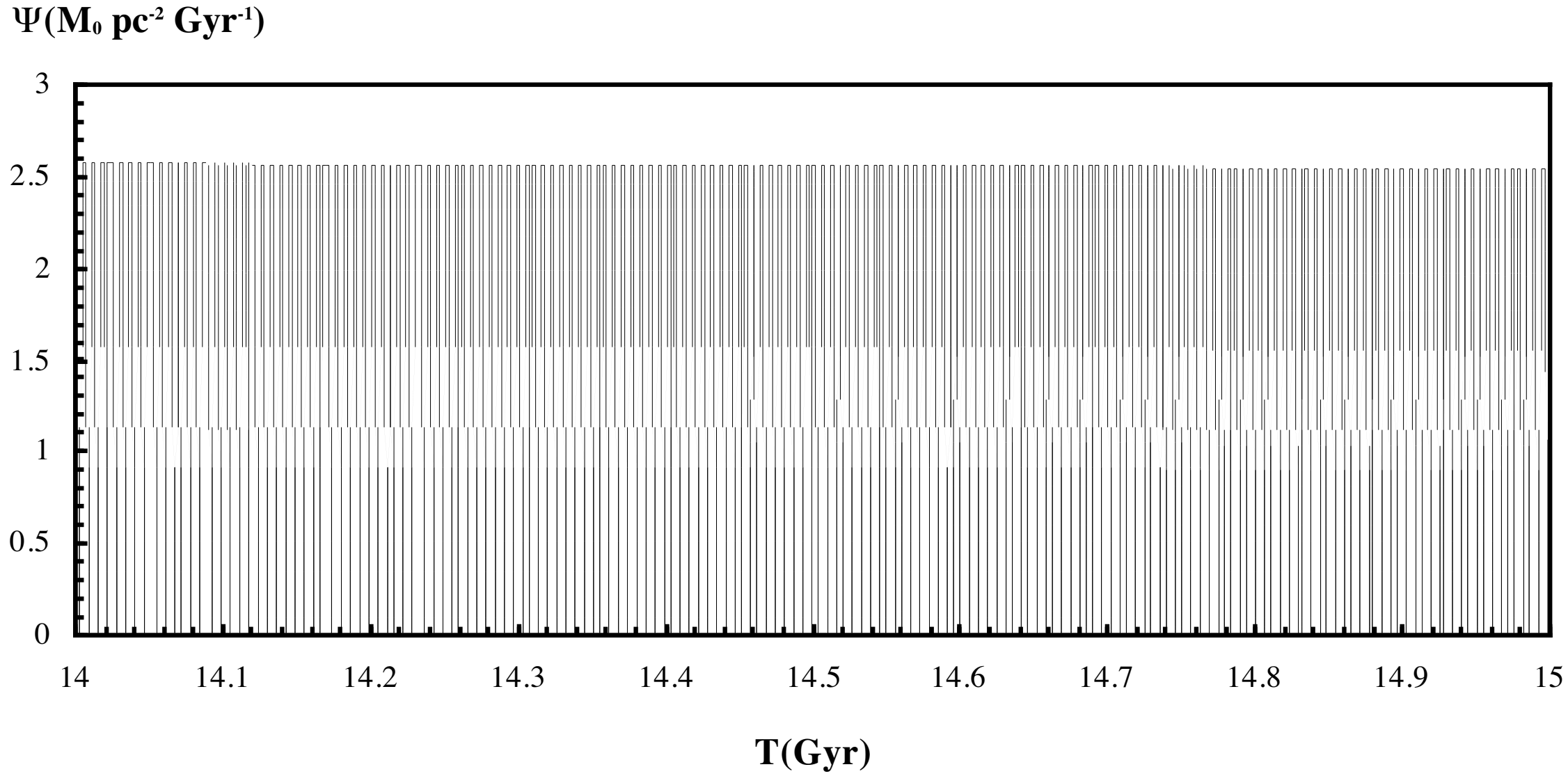

**Figure 2b:** A blow-up of the recent star formation rate of the Solar neighbourhood predicted by Galactic CEMs.

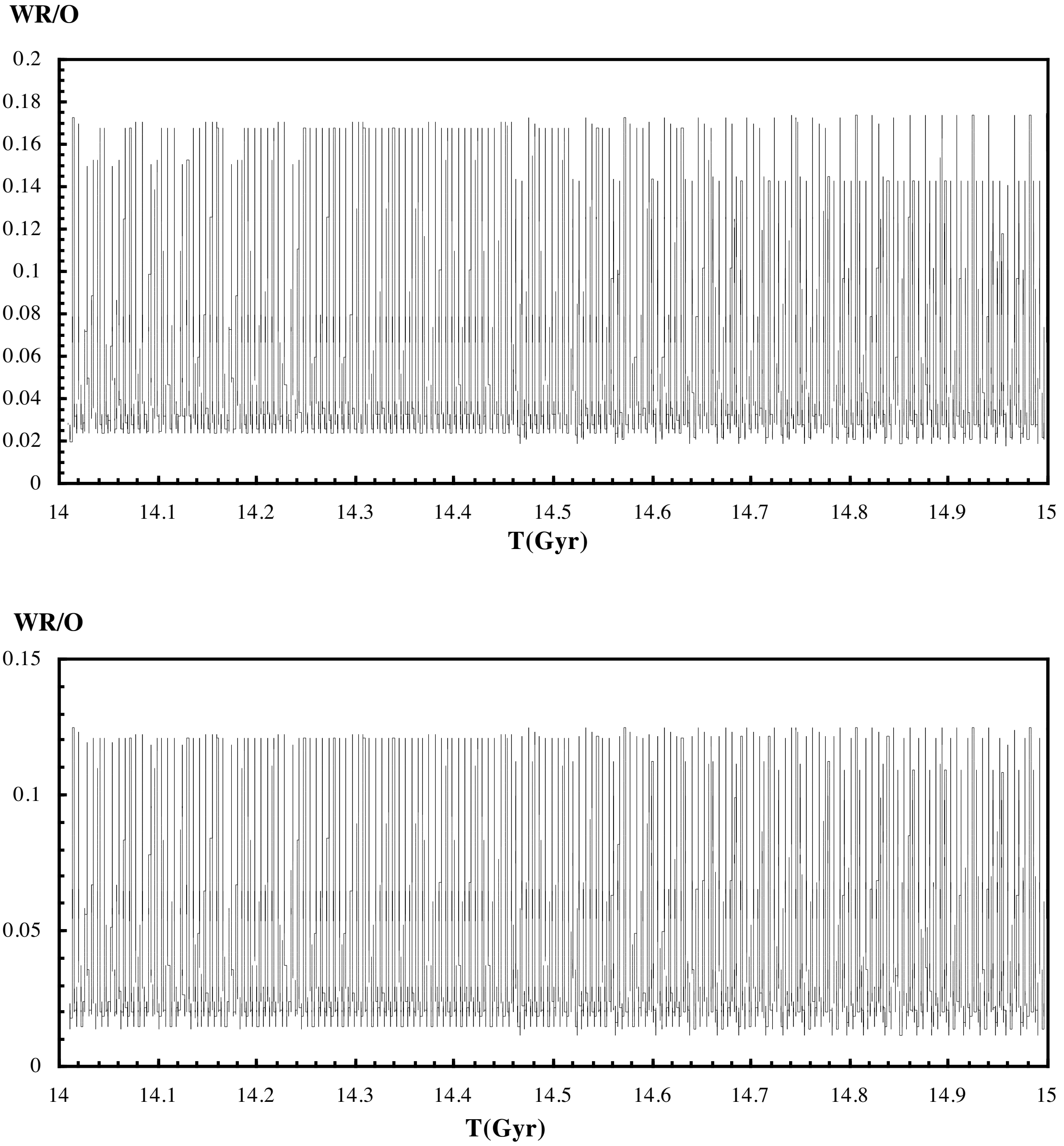

**Figure 3**: The recent temporal evolution of the WR/O number ratio in the Solar neighbourhood. The top (resp. bottom) Figure illustrates simulations assuming a minimum WR mass of 5 $M_o$ (resp. 8 $M_o$).